# Gelation and Re-entrance in Mixtures of Soft Colloids and Linear Polymers of Equal Size

D. Parisi,[1,2,3*] D. Truzzolillo,[4*] A. H. Slim,[5] P. Dieudonné-George,[4] S. Narayanan,[6] J. C. Conrad,[5] V. D. Deepak,[7] M. Gauthier,[7] and D. Vlassopoulos[1,2]

[1]FORTH, Institute of Electronic Structure and Laser, 70013 Heraklion, Crete, Greece

[2]Department of Materials Science and Technology, University of Crete, 70013 Heraklion, Crete, Greece

[3]Department of Chemical Engineering, Product Technology, University of Groningen, Nijenborgh 4, 9747 AG Groningen, The Netherlands

[4]Laboratoire Charles Coulomb (L2C), UMR 5221 CNRS Université de Montpellier, Montpellier, France.

[5]Department of Chemical and Biomolecular Engineering, University of Houston, Houston, Texas 77204-4004, United States

[6]Advanced Photon Source, Argonne National Laboratory, Argonne, IL 60439, United States

[7]Department of Chemistry, University of Waterloo, Waterloo, Ontario N2L 3G1, Canada

*corresponding authors: *domenico.truzzolillo@umontpellier.fr*
*d.parisi@rug.nl*

**ABSTRACT**

Liquid mixtures composed of colloidal particles and much smaller non-adsorbing linear homopolymers can undergo a gelation transition due to polymer-mediated depletion forces. We now show that the addition of linear polymers to suspensions of soft colloids having the same hydrodynamic size yields a liquid-to-gel-to-re-entrant liquid transition. In particular, the dynamic state diagram of 1,4-polybutadiene star–linear polymer mixtures was determined with the help of linear viscoelastic and small angle X-ray scattering experiments. While keeping the star polymers below their nominal overlap concentration, a gel was formed upon increasing the linear polymer content. Further addition of linear chains yielded a re-entrant liquid. This unexpected behavior was rationalized by the interplay of three possible phenomena: (i) depletion interactions, driven by the size disparity between the stars and the polymer length scale which is the mesh size of this entanglement network; (ii) colloidal deswelling due to the increased osmotic pressure exerted onto the stars; and (iii) a concomitant progressive suppression of depletion efficiency on increasing polymer concentration due to reduced mesh size, hence a smaller range of attraction. Our results unveil an exciting new way to tailor the flow of soft colloids and highlight a largely unexplored path to engineer soft colloidal mixtures.

**I. INTRODUCTION**

Colloidal mixtures have emerged as model systems to tailor the phase behavior and flow properties of suspensions, with implications extending from fundamental understanding[1–12] to ubiquitous applications[13–17]. It has been 20 years since the seminal work of Pham *et al.*[18] highlighted the effect of depleting agents on concentrated hard sphere suspensions, and more than 60 years since Asakura and Oosawa[19] unveiled the entropic nature of the attractions between colloids in the presence of smaller particles.

Model hard sphere colloidal suspensions and their mixtures have been intensively studied, whereas the effects of particle softness have been less explored. When a soluble polymer is added to a colloidal suspension, the interactions between colloidal particles are modified and the macroscopic response of the suspension can change dramatically. For example, a transient colloidal network may be formed at low particle volume fractions, below the particle concentration where the dynamics freeze and the glass transition occurs. In this case, gelation (or flocculation) is driven by forces of entropic nature, namely depletion forces. This phenomenon has been observed in a variety of mixtures, with polymers or colloids playing the role of depletants, and hard or soft particles experiencing depletant-mediated attractions[1,5–7,9,11,20–25]. The depletant-to-depleted particle size ratio determines the attraction range and the depletant concentration determines the attraction strength between a pair of particles[19,26]. This scenario holds for colloids whose internal microstructure and shape are not affected by the addition of depleting agents, for example hard spheres. However, for soft colloidal particles, whose microstructure depends on the osmotic pressure of the surrounding medium[27,28], the precise state diagram is currently unknown, and whether a depletion gel state persists upon increasing the depletant concentration (attractions) remains an open question[29].

Star polymers[1,4,5,7–10,30] represent perfect model systems for soft colloids, thanks to their tunable softness dictated by the branching functionality and the degree of polymerization of their arms[21]. In addition, despite the demanding synthesis, they are relatively simple in the sense that they consist of homopolymer arms without enthalpic effects, they can span the entire concentration range from solution to the melt, and their dynamics are reasonably well understood. Recently[31], we addressed the transition from confined-to-bulk dynamics of linear homopolymers added to non-dilute star polymer solutions, with linear and star polymers having nearly the same hydrodynamic size. To date, emphasis was placed on problems involving star polymers well above their overlap concentration, typically in the glassy state. In the present work, by using the same systems as Parisi *et al.*[31], we investigated the dynamic state diagram of star polymers in a good solvent, below their overlap concentration, upon addition of linear polymer chains. We address in particular, three fundamental questions: What is the consequence of adding linear polymer chains to a liquid-like star-polymer suspension at fixed number density and equal hydrodynamic size? How does the interplay between osmotic shrinkage and depletion (mediated by the size of the depletant) control the dynamics of the mixtures? Does the scenario encountered for hard-sphere-polymer mixtures, with a unique liquid-to-gel transition, still hold?

We found that, starting from a solution of star polymers of different softness (functionality) and slightly below their overlap concentration ($C^*$), the addition of linear polymers of equal size drives the system to a dynamic arrest. Such an arrested state is the result of depletion forces exerted on the stars and ruled by a characteristic correlation length of the order of the mesh size of the formed topological chain network. A further increase of the concentration of linear chains promotes a re-entrant liquid state, whose rheological response is mediated by the linear polymer content. This novel re-entrance is further corroborated by the behavior of the mixtures at lower star concentrations. Indeed, the progressive addition of linear polymers yields: i) first a critical gel, which is followed by typical solid-like (network) behavior with detectable structural relaxation, ii) a correlated amorphous liquid (exhibiting liquid-like order) with an observable relaxation dictated by the center-of-mass motion of the stars (colloidal mode), and iii) a viscoelastic liquid whose response is dominated by the contribution of the polymeric network. SAXS experiments supported the scenario of star deswelling, due to the stars themselves (packing effect) and/or the increasing concentration of linear chains (osmotic shrinkage). At the same time, the reduced polymer network mesh size for increasing concentrations indicates that a smaller depletant size leads to a smaller attraction range, corroborating the existence of a (metastable) liquid state due to the progressive suppression of depletion efficiency[32–34].

## II. EXPERIMENTAL
### II.1. Materials

We used two multiarm 1,4-polybutadiene (PBD) stars, identified as S362 and S1114, with number-average branching functionality, weight-average arm molar mass and respective polydispersity equal to $f = 362$ arms, $M_w^a = 24400$ g/mol, $M_w/M_n = 1.06$, and $f = 1114$ arms, $M_w^a = 1270$ g/mol, $M_w/M_n = 1.06$[35]. Both stars were used in previous work, with the high functionality star S1114 being considered as a nearly hard colloidal particle[7,36,37]. Relevant details on the synthesis and the size exclusion chromatography (SEC) analysis of these samples are reported elsewhere[20,35]. According to the well-known Daoud-Cotton model[38], a star polymer in a good solvent is characterized by a non-homogeneous monomer density distribution that comprises three regions: an inner melt-like core, an intermediate ideal region and an outer excluded volume region. The latter is involved in interactions with neighboring stars in crowded suspensions.

Two 1,4-polybutadiene linear polymers, identified as L1000 and L243, were also used. L1000 was obtained from Polymer Source (Canada) and has a weight-average molar mass $M_w$ = 1060000 g/mol and $M_w/M_n$ = 1.1. L243 was provided by Prof. N. Hadjichristidis (KAUST) and has $M_w$ = 243000 g/mol and $M_w/M_n$ < 1.1.

The polymers were dissolved in squalene, a non-volatile solvent providing good (nearly athermal) solvency conditions for 1,4-polybutadiene[39]. The hydrodynamic radius was determined with dynamic light scattering (DLS) measurements under dilute conditions. The hydrodynamic radii ($R_H$) and overlap concentrations ($C^*$) for all the investigated samples are reported in Table 1. The DLS characterization of L243 at 20 °C, illustrated in Fig. S1 of the Supplementary Material, yielded $R_H$ = 15.3 nm and $C_L^*$ = 27 mg/ml. The star–linear polymer mixtures S352/L1000 are characterized by a hydrodynamic size ratio equal to 39/41 = 0.95, whereas the respective mixtures of hard-like spheres and linear polymers S1114/L243 have a ratio of 12/15.3 = 0.78. The nearly identical size of the stars and linear chains, as well as the different star softness, are responsible for the unusual and rich dynamic phase behavior presented in this work. The molecular characteristics of the polymers, together with the light scattering characterization results, and the volume fraction at the glass transition for the stars, $\phi_g$, taken from previous work[7,31,37] are summarized in Table 1. The mixtures were always prepared starting from a pure star polymer solution below $\phi_g$. When linear polymer chains were added, the same star volume fraction $\phi_s = C/C^*$ was maintained, so that the linear chains did in fact replace part of the solvent. To distinguish it from the soft stars, the volume fraction of the hard-like colloids, S1114, will be hereafter identified as $\phi_{HS}$. The linear polymer concentration, in both mixtures and pure solutions, is expressed as the nominal concentration (mg/ml) of chains excluding the stars, i.e,. $C_L = \frac{W_L}{\left(V_{sol} + \frac{W_L}{\rho_L}\right)}$, where $W_L$ and $\rho_L$ are the mass and the density (892 mg/cm$^3$) of the dissolved linear polymer chains and $V_{sol}$ is the volume of the solvent (squalene)[31].

The linear chains in the mixtures were always entangled. The entanglement volume fraction of the linear chains can be estimated as $\phi_e = \left(\frac{M_e}{M_w}\right)^{0.76}$ for good solvents[40], where $M_e$ is the entanglement molar mass (1850 g/mol[41]). This yields $\phi_e = 0.008$ and $\phi_e = 0.24$ for L1000 and L243, respectively. The lowest linear polymer volume fractions ($\phi_L = C_L/C_L^*$, with $C_L$ being the concentration of linear chains) probed in this work were $\phi_L = 3$ and $\phi_L = 1.1$ for L1000 and L243, respectively. The characteristic length of an entangled linear polymer matrix is given by the correlation length $\xi = \left(\frac{kT}{G_N^0}\right)^{1/3}$, with $kT$ being the thermal energy, and $G_N^0(\phi) = G_N^0(\phi = 1)\phi^{2.3}$ ($G_N^0(\phi = 1) = 10^6$ Pa[41]), the diluted polymer plateau modulus[40]. The correlation length $\xi$ can be thought of as the average spatial distance between neighboring entanglement points and is the length dictating the range (and the magnitude) of depletion interactions in semidilute solutions[34,42]. All the experiments were performed at 20 °C.

**Table 1.** Molecular characteristics of the star and linear polymers

| Code | $M_w$ [kg/mol] | $M_w^a$ [kg/mol] | $f$ [-] | $R_H^a$ [nm] | C*(or $C_L^*$) [g/ml] [b.] | $C_g^c$ [g/ml] | $\phi_g = C/C_g$ [-] |
|---|---|---|---|---|---|---|---|
| S1114 | 1415 | 1.27 | 1114 | 12.0±0.5 | 0.326 | 0.26-0.30 | 0.8-0.92 |
| S362 | 8832 | 24.4 | 362 | 39.0±2 | 0.062 | 0.09-0.12 | 1.5-2 |
| L1000 | 1060 | - | - | 41.0±0.8 | 0.006 | - | - |
| L243 | 243 | - | - | 15.3±0.5 | 0.027 | - | - |

[a] Estimated from the diffusion coefficient *D*, measured with DLS in squalene at 20 °C, at a concentration 1 mg/ml, in the dilute regime. [b] The overlap concentration was estimated as $c^* = \dfrac{M_w}{\frac{4}{3}\pi R_H^3 N_A}$. [c] The glass transition was estimated from rheological experiments in the linear viscoelastic regime. A suspension not exhibiting terminal relaxation within the frequency range 0.01 - 100 rad/s was technically considered a glass[1,5,7,31,37].

**II.2. Rheology**

The dynamics of the star–linear polymer mixtures were investigated with rheological measurements, which were performed using a sensitive strain-controlled rheometer (ARES-HR 100FRTN1 from TA, USA). Due to the very limited amounts of samples available, a small home-made cone-and-plate geometry (stainless steel cone with 8 mm diameter, 0.166 rad cone angle) was mostly used. At very low concentrations, a 25 mm stainless steel cone (with angle 0.02 rad) was used to increase the torque signal. The temperature was set to 20.00 ± 0.01 °C and controlled using a Peltier plate with a recirculating water/ethylene glycol bath. During an experimental run, the sample (which had a pasty appearance) was loaded on the rheometer, with special attention to avoid the appearance of bubbles, and a well-defined pre-shear protocol was applied such that each sample was subjected to: (i) a dynamic strain amplitude sweep at fixed frequency (100 rad/s) to determine the linear viscoelastic regime, *i.e.*, where the moduli did not show any detectable dependence on strain amplitude; (ii) a dynamic time sweep at large nonlinear strain amplitude (typically 200%) and low frequency (1 rad/s), to effectively shear-melt (i.e., rejuvenate) the sample, as judged by the time-independent first harmonics $G'(\omega,\gamma_0)$ and $G''(\omega,\gamma_0)$ (this step typically lasted 300 s); (iii) a dynamic time sweep for a (waiting) time $t_w \approx 10^5$ s, which was performed in the linear regime to monitor the time evolution of the moduli to steady state, corresponding to an aged sample; (iv) small-amplitude oscillatory shear (SAOS) tests in the frequency range 100-0.01 rad/s, to probe the linear viscoelastic spectra of the aged samples.

It is worth pointing out that the rejuvenation induced by pre-shearing the samples at amplitudes deeply into the nonlinear regime ($\gamma_0 > 200$ %) erases the accumulated aging. Based on the available experimental evidence, the eventual steady state, typically characterized by a viscoelastic response (G'($\omega$),G"($\omega$)) independent of the waiting time for more than 12 hours[31], did not depend on the details of the rejuvenation protocol (the responses of the aged samples did not show any detectable dependence on the preshear amplitude $\gamma_0 > 200$ % and frequency in the range 1 rad/s $\leq \omega \leq$ 10 rad/s). This makes the adopted procedure robust and it has been widely discussed in many of our previous works[6,20,31,37,43].

Additionally, creep experiments were performed to extend the low-frequency region of the oscillatory response. A stress-controlled rheometer MCR 501 (Anton-Paar, Austria), equipped with a stainless steel cone-plate geometry (8 mm diameter, cone angle of 0.017 rad), was used for creep measurements. The temperature was controlled by a Peltier element that also constituted the lower plate. Different stresses were applied to ensure that the response reflected the linear viscoelastic regime. Creep compliance was then converted into dynamic moduli by means of the nonlinear regularization method proposed by Weese[44] (see the Supporting Material for further information).

It should be noted that when these suspensions were out of equilibrium, they exhibit time-dependent dynamics (aging)[31,45–47] which was taken into account. Typical aging time was between $10^3$ and $10^5$ s, depending on the suspension concentration. The data shown hereafter refer only to aged samples, and the influence of aging will not be discussed further. For $\phi > \phi_g$, the star polymer suspensions behave as viscoelastic solids, with both storage (G') and loss (G") moduli weakly dependent on frequency, G' > G" and G" exhibiting a shallow, broad minimum typical of glassy colloids[47–51].

**II.3. Small angle X-ray scattering (SAXS)**

SAXS measurements were performed with an in-house setup (Montpellier). A high-brightness, low-power X-ray tube, coupled with aspheric multilayer optic (GeniX$^{3D}$ from Xenocs) was employed. It delivers an ultralow divergent beam (0.5 mrad, $\lambda = 0.15418$ nm). Scatterless slits were used to give a clean 0.6mm beam diameter with a flux of 35 Mphotons/s at the sample. We worked in a transmission configuration and scattered intensity was measured using a 2D "Pilatus" 300K pixel detector by Dectris (490×600 pixels) with pixel size (area) of 172×172 µm², at a distance of 1.9 m from the sample loaded in cylindrical quartz capillary tubes (Hilgenberg, 1 mm diameter). SAXS data were collected across a scattering wavevector range $0.07$ nm$^{-1} < q < 0.2$ nm$^{-1}$. The temperature was kept fixed (T= $20.0 \pm 0.1$ ºC) via a recirculating water/ethylene glycol bath. All intensities were corrected by transmission and the empty cell contribution was subtracted.

For the star–linear polymer mixture with $\phi_s = 0.83$ and $C_L = 30$ wt% , the SAXS measurements were performed at the beamline 8-ID-I in the Advanced Photon Source at Argonne National Laboratory (USA). The sample was loaded into a Quartz capillary tube (Charles-Supper, 2 mm inner diameter), and then sealed using wax to prevent solvent evaporation. SAXS data across a wavevector range $0.01$ nm$^{-1} < q < 0.3$ nm$^{-1}$ were collected and corrected according to the background scattering of each sample. A copper block with a Peltier plate was used to control the temperature. The captured scattering intensities were processed using a GUI visualization software package that was developed and provided by APS.

**III. RESULTS AND DISCUSSION**

Before presenting the results obtained with the star–linear mixtures, we report on the structural and rheological features of the (reference) pure star system in a good solvent, at different star concentrations. These experiments were crucial for the subsequent study of the mixtures since they allowed us to locate the liquid-to-glass transition of the stars, as well as to inspect structural changes due to the increase of colloid number density.

In Fig. 1A we report the SAXS intensity I(q) after background (empty cell) subtraction, as a function of the scattering wavevector q for the pure S362 star suspensions at different concentrations $2 \leq \phi_s \leq 5$, all above the glass transition. We observe correlation peaks due to the amorphous dense packing of the stars (liquid-like order). The I(q) peaks shift to larger q values as the concentration is increased, pointing to a reduction of the average distance between the scatterers, here the silicon-rich core of the stars. It is possible to determine the average distance between the cores as $d_{cc} = 2\pi a/q_{max}$, where $q_{max}$ is the scattering wavevector at the first low-q peak of the scattered intensity and a = 1.23 is a numerical prefactor reflecting a structure controlled only by two-body correlation[52]. We found excellent agreement between the measured half distance $d_{cc}/2$ and the star radius computed on the basis of osmotic theory[5,26,30,53] (see Table S1 in the Supplementary Information and the inset in Fig. 1A)[31]. We found that $d_{cc}/2$ exhibits a scaling dependence on star concentration which is compatible with that of concentrated and homogenously distributed scatterers ($d_{cc} \sim \phi_s^{-1/3}$ [54]), not very different from the $\phi_s^{-1/5}$ power-law estimated in our previous work for the star radius [31]. Most remarkably, the very good agreement between the $d_{cc}/2$ and $R_0$ values supports the fact that interpenetration between the stars is very limited and that deswelling due to presence of neighboring stars is more severe than for linear chain solutions, $R(C_L) \approx C_L^{-1/8}$. Since the osmotic pressure increases with functionality $f$ [55,56], the stars are more efficient "osmotic compressors" than linear chains[31].

The rheological spectra including the storage (G′) and loss (G″) moduli as a function of oscillatory frequency ω for pure S362 at various volume fractions ($\phi_s$), from below to well above the glass transition, are shown in Fig. 1B. At $\phi_s$ = 0.9, the S362 suspensions exhibit a response typical for a (viscoelastic) liquid, with G″(ω) >> G′(ω) and terminal frequency scaling of 1 and 2, respectively. For $1.5 < \phi_s < 2.0$ the glass transition takes place, and for $\phi_s > 2.0$ the stars, significantly deformed, attain the so-called jammed glass state (see Chapter 6 of Ref. 27).

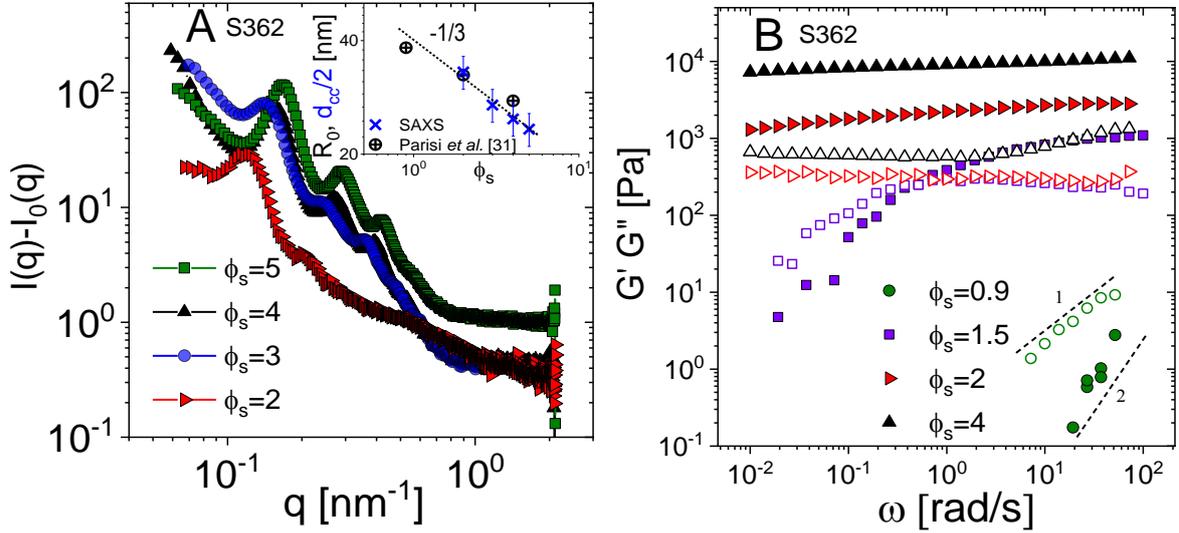

**Figure 1.** (A) SAXS intensity ($I(q)-I_0(q)$), with $I_0(q)$ being the background intensity, for pure S362 star polymer solutions in squalene at various volume fractions $\phi_s$ (see legend) as a function of the scattering wavevector q. Inset: star radius ($R_0$) and average half distance between cores ($d_{cc}/2$) as a function of the star volume fraction ($\phi_s$). The open circles are calculated values from Parisi et al.[31], whereas the X symbols are estimated as $\pi a/q_{max}$ where $q_{max}$ is the scattering wavevector at the low-q intensity peak, and the coefficient a, set to 1.23, reflects a structure controlled only by two-body correlation[52]. The dashed line represents the (-1/3) power-law determined from the SAXS data. (B) Storage modulus $G'$ (solid symbols) and loss modulus $G''$ (open symbols) as a function of the oscillation frequency $\omega$ for pure S362 stars in squalene at various volume fractions $\phi_s$ (see legend) from Parisi et al.[31]. The dashed lines highlight the terminal slopes (see text). All the solutions were measured after steady-state conditions (aging between $10^3$ and $10^5$ s).

In this respect, it is important to specify that here we consider the star deformation to be isotropic and we neglect the possible onset of faceting that may occur at high volume fractions. In addition, as the arm segments in the outer corona region are free to move and rearrange compatibly with their excluded volume, we conjecture that faceting is reduced, if not absent, especially for $\phi_s <1$, where stars preferentially slightly interpenetrate rather than deform. Moreover, all our SAXS results are compatible with the deformation computed assuming an average isotropic compression (Figure 1-A (Inset) and Figure S5 in Supplementary Material).

The addition of linear chains, as detailed hereafter, has a radically different impact on the star suspensions, compared with that of simple star crowding. Fig. 2A shows the LVE response in terms of the frequency-dependent G' and G" for a pure star polymer S362 solution at $\phi_s = 0.9$ and its mixtures with linear chains L1000. The S362 solutions exhibit the typical behavior of a fully relaxed viscoelastic liquid (Fig. 1B). As the linear polymer concentration is increased in the 4.5-5 wt% range, the linear viscoelastic spectra exhibit a liquid-to-solid transition. In this case where the colloid–polymer interactions are purely repulsive, a depleting layer with a thickness proportional to the polymer correlation length $\xi$ forms around each sphere and a depletion attraction occurs between colloids[57]. In other words, the system can be thought of as soft colloids suspended in a sea of uncorrelated polymeric blobs of size $\xi$. Hence, the size ratio that should be considered for depletion effects is the one between the hydrodynamic size of the stars ($R_H = 39$ nm) and the entanglement distance $\xi$ [57]. For instance, if we consider the mixture in Fig. 2A, with L1000 at 5 wt% (circles in Fig. 2A), a correlation length of 16 nm is obtained, corresponding to a size ratio $\xi/R_H \sim 0.4$, which represents a sufficient condition for depletion effects[5,8]. Therefore, a star polymer suspension undergoes gelation upon addition of an entangled network of linear homopolymer chains having nearly the same hydrodynamic radius.

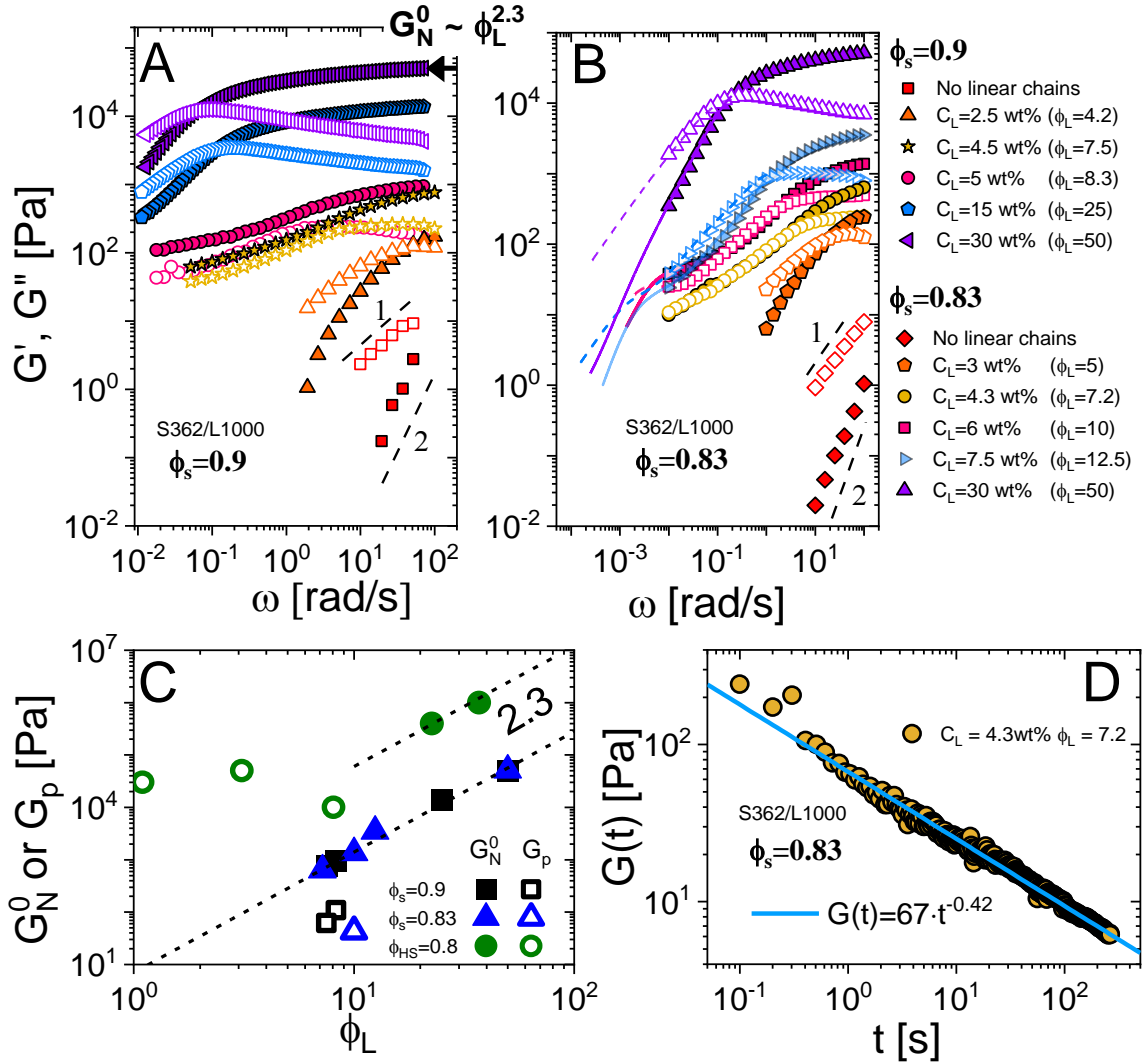

**Figure 2.** Linear viscoelastic spectra in terms of G′ (solid symbols) and G″ (open symbols) as a function of the oscillation frequency ω for S362-L1000 mixtures at various concentrations of L1000 (see legends) and fixed star volume fractions: $\phi_s$ = 0.9 (A) and $\phi_s$ = 0.83 (B). Data at $\phi_s$ = 0.5 and $\phi_s$ = 0.7 are reported in the Supplementary Material (Fig. S4). The horizontal black arrow in panel A indicates the diluted plateau modulus ($G_N^0$) of a pure linear chain solution at 30 wt% (see text). Solid and dashed lines in panel B are creep data converted to dynamic moduli (see Fig. S2 of the Supplementary Material). The dotted lines with slopes 1 and 2 indicate a fully relaxed viscoelastic fluid. C) Polymeric ($G_N^0$) (solid symbols) and colloidal ($G_p$) (open symbols) plateau moduli as a function of the linear chain volume fraction $\phi_L$. The dashed lines

report the power-law dependence of entangled linear polymer chains in a good solvent[40]. D) Stress relaxation modulus G(t), obtained from a step-strain experiment, for $C_L$ = 4.3 wt% as a function of time. The solid line represents the best fit to a power-law function G(t) = *67t^{-0.42}* (see text). The strain amplitude applied for the step-strain test was 1%, well within the linear viscoelastic regime. All the solutions were measured after steady-state conditions (aging between $10^3$ and $10^5$ s).

Quite strikingly, a further increase in linear chain concentration leads to a re-entrant liquid state (see data at $C_L$ = 30 wt% in Fig. 2A). The conditions under which we find such an unexpected multiple transition are different as compared with those reported in previous work on star–linear homopolymer mixtures[5,20]. This is mainly due to two reasons: i) the initial star polymer suspension is a viscoelastic liquid, whereas previously $\phi_s(\phi_L = 0) > \phi_g$, and ii) the hydrodynamic size of the stars and the linear chains is nearly identical. In Ref. 20, the addition of small linear chains (with $\frac{R_H^{linear}}{R_H^{stars}} = 0.05$) to a depleted star polymer glass yielded a re-entrant viscoelastic solid that was identified as a gel, whereas in Ref. 5, the addition of large linear chains to a star glass (with $\frac{R_H^{linear}}{R_H^{stars}} = 0.8$) did not promote any solid-to-liquid transition, even at large linear polymer chain fractions. We also note for completeness that in large hard sphere–small linear polymer mixtures, the attractive glass is established as a re-entrant state resulting from the continuous addition of polymers to the depleted repulsive glass[18,58]. In the present work a gel is promoted by depletion, while the further addition of linear chains results in a re-entrant liquid. We attribute such a re-entrant liquid transition to the continuous action of osmotic forces mediated by the linear polymer chains. Such osmotic forces are responsible for two distinct effects when the concentration of linear chains is increased: i) they cause the deswelling of the stars[31]; ii) they result in a reduced spatial range of depletion attractions because of the mesh size reduction of the semidilute polymers[32–34]. These two effects can be decoupled by using stars with different softness, namely with a different number and length of arms, therefore deswelling degrees, as discussed later in the text.

In soft colloids or mixtures, a hybrid polymeric and colloidal rheological response can be typically detected, as in the case of previous star–linear polymer mixtures [5], grafted nanoparticle melts[36], or diblock copolymer micelles[59]. The polymeric and the colloidal responses are characterized by two different stress relaxation mechanisms that occur on different time scales and proceed hierarchically. At high frequencies, the response is dominated by the polymer matrix and the plateau modulus reflects the polymeric network response, and is estimated by taking into account only the free volume accessible to the chains[40]. Once the linear chains relax, the dynamics are controlled by the colloidal star polymers, whose fingerprint is usually evident in the low-frequency region (colloidal response), typically < 0.1 rad/s. This dual polymeric and colloidal response is also evident in the present investigated systems, see for example the mixture at $C_L$ = 5 wt% in Fig. 2A. However, with increasing linear polymer concentration, *e.g.* at $C_L$ = 30 wt% (Fig. 2A), the rheological response is dominated by the linear polymers, and a crossover between the moduli emerges at low frequencies. The colloidal response is here masked by the polymer dynamics, and barely detectable in the probed frequency range. The mixtures exhibit the features of an ergodic viscoelastic liquid. The separation of the polymeric and colloidal responses is evident when the high-frequency ($G_N^0$) and the low-frequency (colloidal) ($G_p$) plateau moduli, estimated from the relative minima in the loss factor (G"/G'), are plotted against the linear polymer volume fraction (Fig. 2C). The polymeric (high-frequency) plateau modulus follows the expected trend for linear chains in a good solvent, $G_N^0(\phi) \sim \phi_L^{2.3}$ [40], whereas on the contrary, the colloidal (low-frequency) plateau clearly diverges from the linear polymer scaling.

To study this double transition in more detail and confirm the presence of this new re-entrance in the state diagram of equal-size star–linear polymer mixtures, we investigated a slightly lower star polymer concentration ($\phi_s = 0.83$). We also performed complementary creep experiments (see also Fig. S2 in the Supplementary Material) to extend the probed frequency window. A remarkable and even richer behavior of the mixtures is observed upon addition of linear chains at this $\phi_s$ (Fig. 2B). Starting from a viscoelastic liquid in the absence of linear polymer, the system evolves towards a critical gel (percolation threshold)[60] upon increasing $C_L$ (see data at $C_L = 4.3$ wt% in Fig. 2B). Its dynamic moduli G′ and G″ overlap and follow a power-law dependence on frequency over a wide frequency range. Following Ref. 60, we can express the equivalent stress relaxation modulus at $C_L = 4.3$ wt% as $G(t) = S_g t^{-n}$, where $S_g$ is the strength of the gel and n is the relaxation exponent (Fig. 2D). When n assumes a value of 0 or 1, $S_g$ represents the plateau modulus or the viscosity, respectively. In the present case, n was equal to 0.42 and the strength of the gel was 67 Pa s$^{0.42}$. Although there are no universal values for n and $S_g$, due to the nature of the gelling system investigated, our values are comparable to those reported in the literature for critical colloidal gels ($0.4 < n < 1$)[60–62].

The emergence of such criticality is universal when gelation is induced in an initially well-dispersed colloidal liquid, and it represents very important evidence for the progressive buildup of the gel structure for increasing $C_L$. In contrast, this is the first time to our knowledge that such critical behavior is observed at such a high colloidal volume fraction ($\phi_s = 0.83$): colloidal softness hampers gel formation in the range of volume fractions ($\phi_{gel}$) where gels of hard particles typically form, from a few percent up to the vitrification threshold $\phi_{gel} \cong 0.58\text{-}0.60$[63].

A further increase in $C_L$ drives the system to solid-like behavior at low frequencies, with the emergence of a low-frequency colloidal plateau modulus (see data at $C_L = 6$ wt% in Fig. 2B).

The latter assumes a value of about 40 Pa, yielding an apparent colloidal correlation length $\xi_c = \left(\frac{kT}{G_N^0}\right)^{1/3}$ = 48 nm, slightly larger than the colloidal radius ($R_H$ = 39 nm). This suggests a non-interpenetrating condition for the stars. It is worth noting that for star polymer glasses (in the absence of linear chains), where colloids are arranged in a cage-like fashion, $\xi_c$ is typically only a fraction of the star radius and has been interpreted as the extent of the overlapping region between two neighboring stars[64]. It is important to stress that the quantity $\xi_c$ should not be confused with the mesh size of the polymer network $\xi$.

To further exclude the presence of caging we can exploit an alternative, yet equivalent, interpretation of the colloidal correlation length based on the analysis of glassy microgel suspensions by Cloitre *et al.*[65], where the plateau modulus is linked to the maximum displacement of a generic particle with radius $R_H$ inside its topological cage. When a soft particle with radius $R_H$ moves over a distance $\delta$, it deforms elastically its neighbors, which in turn push it back inside the cage. The restoring energy that drives the particle back can be written as $G_N^0 \delta^2 R_H$. When the latter equals the thermal energy, a maximum displacement $\delta$ is reached. This simple model was applied to the present star–linear polymer mixtures to estimate the maximum displacement of the constrained stars in the polymer matrix, yielding $\delta$ = 50 nm, *i.e.,* larger than the star hydrodynamic radius. By combining the two findings, $\xi_c$ = 48 nm and $\delta$ = 50 nm, it is possible to assert that stars are not truly in a glassy state, but rather in a percolated network with reduced possibility to diffuse, due to the presence of the polymeric network and depletion attractions. We recall that in repulsive glasses, according to the Lindemann criterion[64,66], the maximum displacement is typically at most one tenth of the particle radius (about 4 nm here). We further recall that the pure star polymer solution at this concentration ($\phi_s = 0.83$) exhibits liquid-like behavior (Fig. 2B). We conclude that caging must be excluded.

As already witnessed for the star–linear polymer mixtures at $\phi_s = 0.9$, the progressive addition of linear chains to mixtures with $\phi_s = 0.83$ gives rise to a re-entrant liquid (see data at $C_L = 7.5$ wt% in Fig. 2B). However, the viscoelastic character of the mixtures and the role of the stars as outlined above is also evident here, since the terminal flow regime is not attained over the examined frequency range. Indeed, the power-law behavior of the moduli ($G' \sim \omega^2$, $G'' \sim \omega^1$) typical for a fully relaxed liquid is not observed even at the highest concentration investigated here ($C_L = 30$ wt%). Overall, whereas for mixtures with $C_L$ below 7.5 wt% the rheological response is significantly affected by the presence of stars, at higher values of $C_L$ the linear polymer chains dominate the linear viscoelastic spectra, however without fully suppressing the colloidal response.

We also remark that, unlike the case described in Fig. 2A ($\phi_s = 0.9$), where a dynamically arrested state is clearly attained over several decades in frequency upon addition of linear chains, for a slightly lower value of $\phi_s = 0.83$, the concentration of star polymers is not high enough to promote a long-lasting arrested state. In fact, investigations at even lower star polymer volume fractions ($\phi_s = 0.5$, and $\phi_s = 0.7$) displayed no dynamic arrest at any $C_L$ (see Fig. S4 in the Supplementary Material). Nevertheless, the colloidal mode at low frequencies was still observable.

To further support the depletion-gel picture, exclude the possibility of chain-mediated bridging of stars, explore the influence of star polymer softness, and decouple the dual deswelling-mesh size reduction effect, we also investigated nearly symmetric, in hydrodynamic size (see Table 1), S1114/L243 star–linear polymer mixtures. Here, the soft star was replaced with a hard-sphere-like (HS) star (S1114).

The frequency-dependent G' and G" of S1114/L243 mixtures with a fixed volume fraction $\phi_{HS}$ (below $\phi_g$) of S1114 and increasing fractions of L243 is depicted in Fig. 3. The dynamic frequency spectrum of the L243 melt is also included in the figure for comparison. The viscoelastic liquid S1114 star at $\phi_{HS} = 0.8$ undergoes gelation upon addition of 3 wt% of L243. Note that at 3 wt% the L243 chains are entangled (see Materials section). This suggests again that the pivotal length scales for depletion are the size of the stars and the average correlation length (mesh size) of the polymer network, as also observed by Verma *et al.*[57].

As the concentration of linear chains is increased the S1114 particles, which still conserve the deformability of soft colloids[7], deswell due to the osmotic pressure exerted by the linear chains. This is reflected in the significant reduction in the low-frequency moduli at $C_L = 22$ wt% of L243, and the remarkable separation between the polymeric and colloidal responses in the investigated frequency range[5,36]. However, a further increase of $C_L$ to 61 wt% does not lead to a fully relaxed viscoelastic liquid, as contrarily attained at $C_L = 100$ wt%. The hard-sphere-like stars still manifest their presence with the onset of a slow mode, which becomes progressively weaker in favor of terminal relaxation dynamics dominated by the linear chains[7].

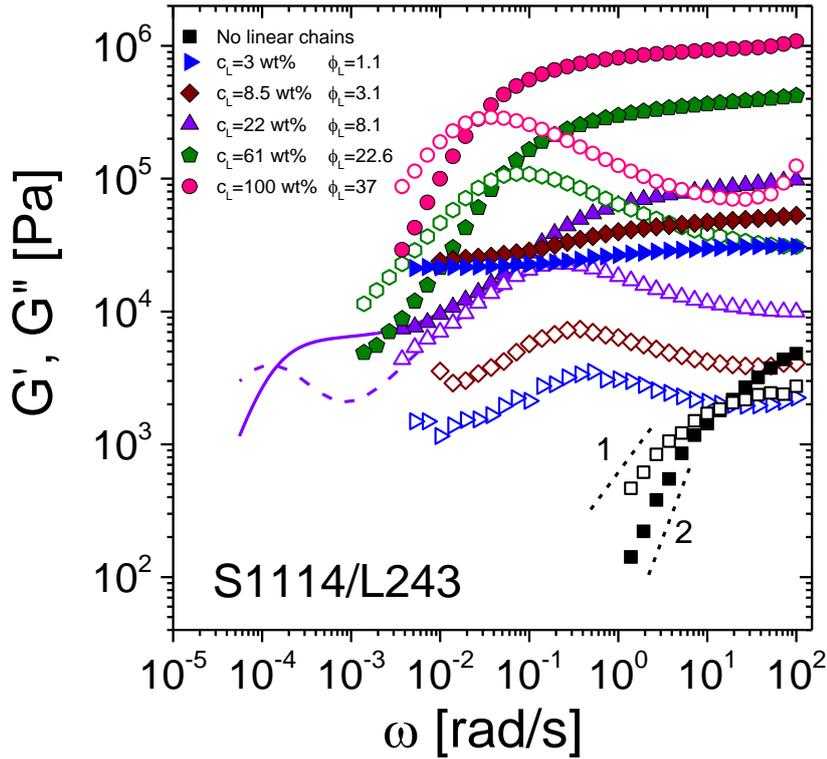

**Figure 3.** Linear viscoelastic spectra in terms of G′ (solid symbols) and G″ (open symbols) as a function of the oscillation frequency ω for the S1114–L243 mixtures. The squares represent the hard-sphere-like star polymer (S1114) suspension at $\phi_{HS} = 0.8$, below the glass transition (Table 1). The right-pointing triangles, diamonds, up-pointing triangles, pentagons, and circles correspond to mixtures at $C_L$ = 3.0, 8.5, 22, 61, and 100 wt%, respectively. The circles correspond to pure linear chains. The lines are from creep data converted to dynamic moduli. The short-dashed lines highlight the terminal slope of fully relaxed viscoelastic fluids. All the solutions were measured after steady-state conditions (aging between $10^3$ and $10^5$ s).

Therefore, the phenomenology encountered for the HS-like star–linear mixtures (S1114–L243) displays the same general features as the ones characterizing the softer star suspensions (S362–L1000), with a solid (gel) pocket of states separating the two seas of liquid states. Since the overall scenario remains unchanged, we speculate that an important role is played by the loss of efficiency of the depletion mechanism at very high $C_L$ (due to the decrease in mesh size ξ), since this would intervene even in model HS systems.

To highlight more clearly the role of softness, we can proceed one step further and directly compare our data to determine whether the different star functionalities quantitatively tunes such a reentrant behavior. To this end we compiled the rheological results discussed above in the form of a state diagram, which is depicted in Fig. 4 in terms of linear chains volume fraction ($\phi_L$) and volume fraction of the stars ($\phi_s$). Rheological results for the mixtures at $\phi_s = 0.5$ and $\phi_s = 0.7$ are reported in the Supplementary Material (Fig. S4). The assignment of solid-like behavior is based on the linear viscoelastic spectra. In the diagram we consider as viscoelastic solids those suspensions showing a storage modulus exceeding the loss modulus, at least down to 0.01 rad/s, meaning a structural relaxation time exceeding 100 s[6,20,31,37,43]. An exception is represented by the mixture at $\phi_s = 0.83$ and $C_L = 4.3$ wt%, *i.e.*, when critical gel behavior was observed.

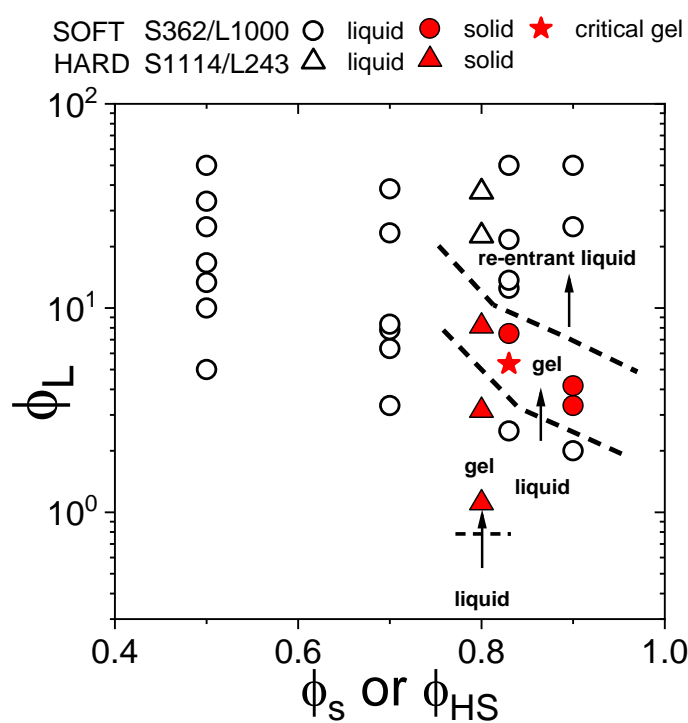

**Figure 4.** State diagram for equal-size star–linear polymer mixtures: S362–L1000 (circles and stars) and S1114–L243 (triangles) mixtures in the $\phi_L$–$\phi_s$ plane. The open symbols, filled star and filled circles indicate the liquid, critical gel and solid states, respectively. The arrows and the dashed lines indicate the approximate location of the boundaries of the liquid-to-solid and the solid-to-re-entrant liquid transitions. It is worth recalling that $\phi_g$ is in the range 0.8-0.92 for S1114 and 1.5-2 for S362 (see Table 1).

As can be seen in Fig. 4, the two sets of mixtures (S362–L1000, S1114–L243) share the same qualitative behavior, but the extent of the solid-like pocket can apparently be tuned by the colloidal softness: the gel pocket is wider in the case of S1114–L243 mixtures, even though in this case the star number density is slightly lower than that characterizing the softer S362–L1000 systems, for which re-entrance is observed. We speculate that enhanced osmotic deswelling in the S362–L1000 mixtures reduces the extent of the solid-like behavior (see Fig. 4), favoring the emergence of the re-entrant liquid at lower $\phi_L$. Conversely, depletion forces appear to be more effective in hard star suspensions[1] that undergo gelation at much lower $\phi_L$, compared to the soft stars.

Hence, our rheological data shed light on a new aspect of the state diagram of soft colloids in the presence of depleting agents forming networks: The osmotic shrinkage of colloids due to the presence of linear chains, and the loss of depletion efficacy together give rise to a re-entrant liquid. We have shown that the addition of polymeric depletants has a dual effect on a liquid soft colloidal system with a fixed particle number density. Depletion attractions and osmotic deswelling impact the structural relaxation of the suspensions: the first contribution induces gel formation, while the second one leads to its melting.

At the same time, star polymer deswelling causes an increase in free volume, which facilitates the relaxation of linear chains. The SAXS data confirm this scenario and provide evidence for the central role of star deswelling, as demonstrated in Fig. 5, which depicts the SAXS intensity I(q) as a function of the scattering wavevector for two star–linear polymer mixtures at various linear chain concentrations. The volume fractions of the pure star polymer solutions are $\phi_s$ = 0.83 and $\phi_s$ = 4. The limited data set obtained at low star polymer volume fractions reflects the lack of material available and a weak scattering signal. Nevertheless, two important messages can be drawn from this figure: (i) For the mixture at $\phi_s$ = 4, the structural peaks shift towards larger scattering wavevector as the concentration of linear chains is increased. This horizontal shift is remarkable at $C_L$ = 30 wt% (see lozenges in Fig. 5) and implies that the distance between scatterers (mainly the silicon-rich star cores, whose number density is fixed) decreases as $C_L$ is increased. The low-q intensity peak gives access to the average distance between star cores, and if we consider that the stars are in contact due to depletion forces and that interpenetration is scarce (as expected for high-$f$ stars in a good solvent[7]), a smaller effective radius can be computed for the (deswollen) colloidal stars as R = $\pi a/q_{max}$. This is shown in the inset of Fig. 5. Here we may note that at $C_L$ = 30 wt% and $\phi_s$ = 4, the size of the stars (18 nm) is not far from that corresponding to their calculated collapsed state (16 nm)[7,31,38] and, most importantly, it is in excellent agreement with the star size computed from free volume considerations as reported in Ref. 31 and the osmotic theory (see Supplementary Material). (ii) Shrinkage of the stars is also evident at lower number densities, as seen in Fig. 5 for $\phi_s$ = 0.83 and $C_L$ = 30 wt%. The effective star radius obtained from the observed correlation peak is about 21 nm, and by considering $R_H$ = 39 nm (Table 1) it is possible to estimate a star shrinkage of nearly 50%. We recall that the rheology of the mixture at $\phi_s$ = 0.83 and $C_L$ = 30 wt% (see Fig. 2) clearly displayed viscoelastic liquid behavior with a barely detectable colloidal mode. Interestingly, the net attraction between stars arising from unbalanced osmotic forces is confirmed by the low-q power-law upturn I(q) ~ $q^{-3.3\pm0.1}$ in the SAXS spectrum for the $\phi_s$ = 0.83 and $C_L$ = 30 wt%. We probed the surface fractal region [67,68] in Fourier space, where I(q) ~ $q^{-(6-d_s)}$, from which we obtained a surface fractal dimension $d_s$ = 2.7±0.1. In gels, $d_s$ is a measure of the roughness between the colloid-rich and the colloid-poor regions, ranging between 2 (smooth surfaces) and 3 (infinitely rough surfaces). The obtained value therefore suggests the presence of very rough interfaces and the formation of highly porous structures. Consequently, our results indicate that these depletion gels of star polymers do not form bicontinuous networks with sharp interfaces between colloid-rich and colloid-poor regions. This is in sharp contrast with the gel structure of hard-sphere–polymer mixtures, in which the Porod scaling $I(q) \sim q^{-4}$, corresponding to $d_s = 2$, has been observed most commonly[69].

Last but not least, we compared our shrinkage results from SAXS with the theoretical predictions based on Flory-type arguments[5,26,30,53], accounting for the osmotic, elastic and interaction free energy (see discussion and Fig. S5 in the Supplementary Material), obtaining very good agreement (within 10%).

This finding supports the evaluation of the star size via the analysis of the SAXS data, and corroborates the existence of star polymers in close contact being at the origin of the gel-like response probed by linear rheology. A short note on the reduction of the effective volume fraction of the stars follows. Based on the values of the radii reported in the inset of Fig. 5 the effective volume fraction decreases from $\phi_s = 4$ (at $\phi_L = 0$) to $\phi_s = 0.46$ (at $\phi_L = 50$). For the mixtures undergoing multiple transitions (Fig. 2) the radius of the stars can be inferred from the viscoelastic spectra following the "chemical approach" (see the Supporting Material) described in Ref. 20. We obtain volume fractions ranging respectively from $\phi_s = 0.83$ and $\phi_s = 0.9$ at $\phi_L = 0$ to $\phi_s = 0.15$ and $\phi_s = 0.165$ at $\phi_L = 50$. Hence, it is not surprising that the rheology of the mixtures is highly affected by a drastic reduction of the volume effectively occupied by the colloidal component of the liquid (the stars), causing a drop of the stress response originating from the star-polymer network.

In closing, we emphasize again that the above analysis does not exclude the possible parallel action of the mechanism of suppression of depletion attraction, due to a smaller polymer mesh size for increasing concentrations, as already discussed.

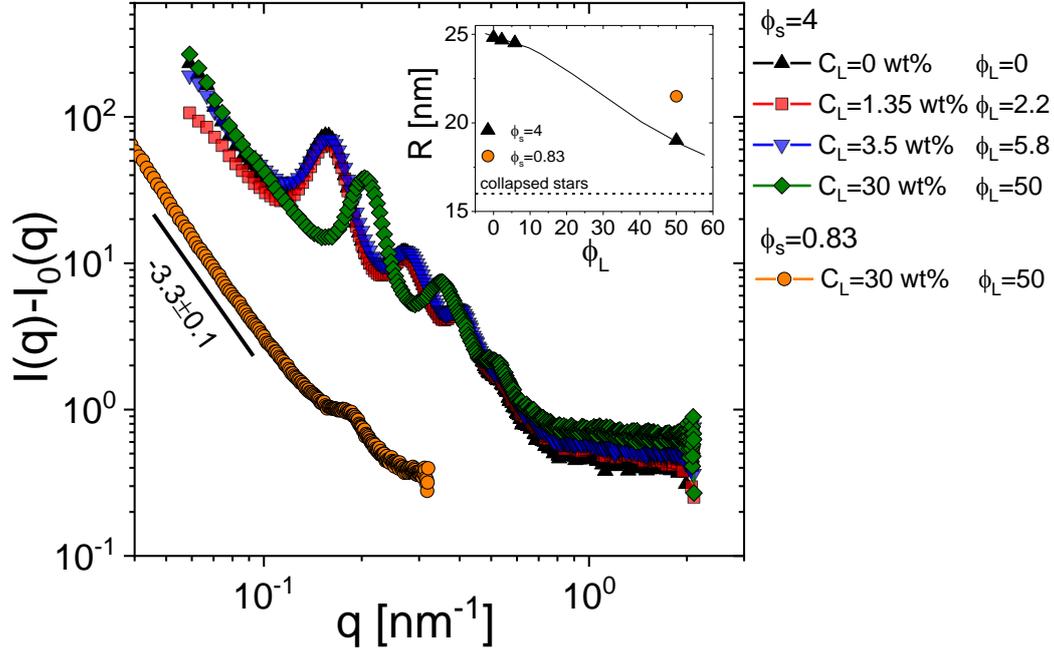

**Figure 5.** SAXS intensity (I(q)-I$_0$(q)) for the S362–L1000 mixtures in squalene at fixed $\phi_s$ = 0.83 and $\phi_s$ = 4 and various linear polymer concentrations C$_L$ (see legend) as a function of the scattering wavevector q. The line through the $\phi_s$ = 0.83 data represents the best fit obtained with a power-law having an exponent of -3.3. This gives a surface fractal dimension equal to d$_s$ = 2.7±0.1 (see main text). Inset: star radius (R) at $\phi_s$ = 0.83 and $\phi_s$ = 4 as a function of linear polymer volume fraction ($\phi_L$). R was estimated as πa/q$_{max}$ where q$_{max}$ is the scattering wavevector at the low-q intensity peak, and the coefficient a = 1.23 reflects a structure controlled only by two-body correlation[52]. The horizontal dashed line represents the calculated radius of the totally collapsed S362 star[31], whereas the continuous line is a guide for the eye.

## CONCLUSIONS

In this work we have shown that star polymer suspensions (used as paradigms for soft colloids), at concentrations below their overlap concentration, can undergo a liquid-to-gel transition upon addition of entangled linear homopolymer chains having nearly the same hydrodynamic size.

The origin of gelation is attributed, as in the case of hard sphere–linear polymer mixtures, to depletion forces arising from the asymmetry between the size of the star polymer and the correlation length of the linear chain matrix. Remarkably, a further increase in the concentration of linear chains gives rise to an unprecedented re-entrant liquid. This phenomenon is promoted by the gain in free volume due to the deswelling of the stars (osmotic shrinkage) for increasing linear chain concentrations (acting as osmotic compressors). The shrinkage effect is also supported by structural results obtained from SAXS experiments. Indeed, a clear shift observed in the main peaks of the structure factor of star-polymer suspensions, in the presence of linear additives, reflects star deswelling. A parallel effect contributing to the observed re-entrance is the fact that the reduced mesh size at higher polymer concentrations leads to a smaller range of depletion attractions, hence making it less probable.

As the concentration of star polymers is reduced to $\phi_s = 0.83$, further departing from nominal overlap, the sensitivity of the rheological response to the addition of linear polymer increases. A critical-gel condition precedes a solid-like response at low frequencies (solid-like pocket), and a correlated viscoelastic liquid with colloidal response is detected. In such mixtures, the concentration of soft colloids is not sufficiently large to promote an arrested state extending over several decades of frequency, contrary to what was observed at $\phi_s = 0.9$. The rheological data were used to construct a dynamic state diagram of polymer vs star concentration for equal-size star–linear polymer mixtures, which significantly differs from those of respective mixtures with large size asymmetry[5], or star–star[8,9], star–hard sphere[1,7], and hard-sphere–hard sphere[11] mixtures investigated so far. None of these mixtures, all with large size asymmetries, displayed re-entrant melting of a dynamically arrested state. The unique phenomenology that we encountered in the present work largely depends on the depletant-to-colloid size ratio, diffusion properties of the depletants, colloidal softness, and solvency conditions. Contrarily to what we have reported here, largely asymmetric star-linear polymer mixtures can experience a completely different gel-to-liquid-to-gel multiple transition that has been previously discussed [20]. Our volume fraction estimation for the stars further suggests that the reduction in size of soft colloids due to osmotic compression is very relevant for the rheology of mixtures and it emphasizes once again the importance of determining the volume fraction of soft colloids especially in the presence of osmotic compressors.

We conjecture that moving from good to poor solvent condition would cause the linear chains to adsorb on the stars, suppressing depletion and possibly driving a gelation mediated by linear chains, the latter causing energetic bridging between the colloids (the stars here). Both the species (linear and stars) would experience solvophobic attractions when in close contact with another suspended component of the mixtures: star-star, linear-linear and star-linear interactions would include an attractive term that could bring to microphase separation and progressively stabilize the gel phase. Hence, we would expect a totally different outcome, that could be corroborated in future studies.

Blending colloidal systems with different softness levels and molecular structures revealed a simple, yet elegant way to control phase transitions, as well as the flow properties of suspensions.

**Declaration of Competing Interest**

The authors declare to have no competing financial interests or personal relationships that could have influenced the work reported in this paper.

**Acknowledgements**

We would like to thank Professor Nikos Hadjichristidis (KAUST) for providing a linear polymer used in this work and Dr. Julian Oberdisse for fruitful discussions. This research was partly funded by the EU (European Training Network Colldense (H2020-MCSA-ITN-2014, grant number 642774). AS and JCC acknowledge support from the Welch Foundation (E-1869)

**Supplementary Material**

Dynamic light scattering analysis results for sample L243; creep compliance and conversion into dynamic moduli; shear rheology of star (S362)–linear (L1000) mixtures at $\phi_s = 0.5$ and $\phi_s = 0.7$ at various linear chain concentrations; Flory's prediction on the shrinkage of a star polymer in a solution of linear homopolymer chains.

**REFERENCES**

(1) Marzi, D.; Capone, B.; Marakis, J.; Merola, M. C.; Truzzolillo, D.; Cipelletti, L.; Moingeon, F.; Gauthier, M.; Vlassopoulos, D.; Likos, C. N. Depletion, Melting and Reentrant Solidification in Mixtures of Soft and Hard Colloids. *Soft matter* **2015**, *11* (42), 8296–8312.

(2) Stiakakis, E.; Vlassopoulos, D.; Likos, C. N.; Roovers, J.; Meier, G. Polymer-Mediated Melting in Ultrasoft Colloidal Gels. *Physical review letters* **2002**, *89* (20), 208302.

(3) Stiakakis, E.; Petekidis, G.; Vlassopoulos, D.; Likos, C. N.; Iatrou, H.; Hadjichristidis, N.; Roovers, J. Depletion and Cluster Formation in Soft Colloid-Polymer Mixtures. *EPL (Europhysics Letters)* **2005**, *72* (4), 664.

(4) Truzzolillo, D.; Vlassopoulos, D.; Gauthier, M. Rheological Detection of Caging and Solid–Liquid Transitions in Soft Colloid–Polymer Mixtures. *Journal of Non-Newtonian Fluid Mechanics* **2013**, *193*, 11–20.

(5) Truzzolillo, D.; Vlassopoulos, D.; Gauthier, M. Osmotic Interactions, Rheology, and Arrested Phase Separation of Star–Linear Polymer Mixtures. *Macromolecules* **2011**, *44* (12), 5043–5052.

(6) Truzzolillo, D.; Marzi, D.; Marakis, J.; Capone, B.; Camargo, M.; Munam, A.; Moingeon, F.; Gauthier, M.; Likos, C. N.; Vlassopoulos, D. Glassy States in Asymmetric Mixtures of Soft and Hard Colloids. *Physical review letters* **2013**, *111* (20), 208301.

(7) Merola, M. C.; Parisi, D.; Truzzolillo, D.; Vlassopoulos, D.; Deepak, V. D.; Gauthier, M. Asymmetric Soft-Hard Colloidal Mixtures: Osmotic Effects, Glassy States and Rheology. *Journal of Rheology* **2018**, *62* (1), 63–79.

(8) Zaccarelli, E.; Mayer, C.; Asteriadi, A.; Likos, C. N.; Sciortino, F.; Roovers, J.; Iatrou, H.; Hadjichristidis, N.; Tartaglia, P.; Löwen, H. Tailoring the Flow of Soft Glasses by Soft Additives. *Physical review letters* **2005**, *95* (26), 268301.

(9) Mayer, C.; Zaccarelli, E.; Stiakakis, E.; Likos, C. N.; Sciortino, F.; Munam, A.; Gauthier, M.; Hadjichristidis, N.; Iatrou, H.; Tartaglia, P. Asymmetric Caging in Soft Colloidal Mixtures. *Nature materials* **2008**, *7* (10), 780.

(10) Mayer, C.; Stiakakis, E.; Zaccarelli, E.; Likos, C. N.; Sciortino, F.; Tartaglia, P.; Löwen, H.; Vlassopoulos, D. Rheological Transitions in Asymmetric Colloidal Star Mixtures. *Rheologica Acta* **2007**, *46* (5), 611–619. https://doi.org/10.1007/s00397-006-0145-8.

(11) Imhof, A.; Dhont, J. K. G. Experimental Phase Diagram of a Binary Colloidal Hard-Sphere Mixture with a Large Size Ratio. *Physical Review Letters* **1995**, *75* (8), 1662–1665.

(12) Laurati, M.; Petekidis, G.; Koumakis, N.; Cardinaux, F.; Schofield, A. B.; Brader, J. M.; Fuchs, M.; Egelhaaf, S. U. Structure, Dynamics, and Rheology of Colloid-Polymer Mixtures: From Liquids to Gels. *The Journal of chemical physics* **2009**, *130* (13), 134907.

(13) Quirk, R. P.; Kim, J. Recent Advances in Thermoplastic Elastomer Synthesis. *Rubber chemistry and technology* **1991**, *64* (3), 450–468.

(14) Hadjichristidis, N.; Pitsikalis, M.; Pispas, S.; Iatrou, H. Polymers with Complex Architecture by Living Anionic Polymerization. *Chemical reviews* **2001**, *101* (12), 3747–3792.

(15) Wu, W.; Wang, W.; Li, J. Star Polymers: Advances in Biomedical Applications. *Progress in Polymer Science* **2015**, *46*, 55–85.

(16) Ren, J. M.; McKenzie, T. G.; Fu, Q.; Wong, E. H.; Xu, J.; An, Z.; Shanmugam, S.; Davis, T. P.; Boyer, C.; Qiao, G. G. Star Polymers. *Chemical reviews* **2016**, *116* (12), 6743–6836.

(17) Nicolai, T. Gelation of Food Protein-Protein Mixtures. *Advances in colloid and interface science* **2019**, *270*, 147–164.

(18) Pham, K. N.; Puertas, A. M.; Bergenholtz, J.; Egelhaaf, S. U.; Moussaıd, A.; Pusey, P. N.; Schofield, A. B.; Cates, M. E.; Fuchs, M.; Poon, W. C. Multiple Glassy States in a Simple Model System. *Science* **2002**, *296* (5565), 104–106.

(19) Asakura, S.; Oosawa, F. On Interaction between Two Bodies Immersed in a Solution of Macromolecules. *The Journal of chemical physics* **1954**, *22* (7), 1255–1256.

(20) Truzzolillo, D.; Vlassopoulos, D.; Munam, A.; Gauthier, M. Depletion Gels from Dense Soft Colloids: Rheology and Thermoreversible Melting. *Journal of Rheology* **2014**, *58* (5), 1441–1462.

(21) Vlassopoulos, D.; Cloitre, M. Tunable Rheology of Dense Soft Deformable Colloids. *Current opinion in colloid & interface science* **2014**, *19* (6), 561–574.

(22) Mayer, C.; Sciortino, F.; Likos, C. N.; Tartaglia, P.; Löwen, H.; Zaccarelli, E. Multiple Glass Transitions in Star Polymer Mixtures: Insights from Theory and Simulations. *Macromolecules* **2009**, *42* (1), 423–434.

(23) Moreno, A. J.; Colmenero, J. Relaxation Scenarios in a Mixture of Large and Small Spheres: Dependence on the Size Disparity. *The Journal of chemical physics* **2006**, *125* (16), 164507.

(24) Foffi, G.; Götze, W.; Sciortino, F.; Tartaglia, P.; Voigtmann, T. Mixing Effects for the Structural Relaxation in Binary Hard-Sphere Liquids. *Physical review letters* **2003**, *91* (8), 085701.

(25) Germain, P.; Amokrane, S. Equilibrium Route to Colloidal Gelation: Mixtures of Hard-Sphere-like Colloids. *Physical review letters* **2009**, *102* (5), 058301.

(26) Likos, C. N. Effective Interactions in Soft Condensed Matter Physics. *Physics Reports* **2001**, *348* (4–5), 267–439.

(27) Wagner, N. J.; Mewis, J. *Theory and Applications of Colloidal Suspension Rheology*; Cambridge University Press, 2021.

(28) Scotti, A.; Schulte, M. F.; Lopez, C. G.; Crassous, J. J.; Bochenek, S.; Richtering, W. How Softness Matters in Soft Nanogels and Nanogel Assemblies. *Chemical Reviews* **2022**, *122*(13), 11675-11700.

(29) Yuan, G.; Cheng, H.; Han, C. C. The Glass Formation of a Repulsive System with Also a Short Range Attractive Potential: A Re-Interpretation of the Free Volume Theory. *Polymer* **2017**, *131*, 272–286.

(30) Wilk, A.; Huißmann, S.; Stiakakis, E.; Kohlbrecher, J.; Vlassopoulos, D.; Likos, C. N.; Meier, G.; Dhont, J. K. G.; Petekidis, G.; Vavrin, R. Osmotic Shrinkage in Star/Linear Polymer Mixtures. *The European Physical Journal E* **2010**, *32* (2), 127–134.

(31) Parisi, D.; Truzzolillo, D.; Deepak, V. D.; Gauthier, M.; Vlassopoulos, D. Transition from Confined to Bulk Dynamics in Symmetric Star–Linear Polymer Mixtures. *Macromolecules* **2019**, *52* (15), 5872–5883.

(32) Li-In-On, F. K. R.; Vincent, B.; Waite, F. A. Stability of Sterically Stabilized Dispersions at High Polymer Concentrations; ACS Publications, 1975.

(33) Feigin, R. I.; Napper, D. H. Depletion Stabilization and Depletion Flocculation. *Journal of Colloid and Interface Science* **1980**, *75* (2), 525–541.

(34) García, Á. G.; Nagelkerke, M. M.; Tuinier, R.; Vis, M. Polymer-Mediated Colloidal Stability: On the Transition between Adsorption and Depletion. *Advances in colloid and interface science* **2020**, *275*, 102077.

(35) Gauthier, M.; Munam, A. Synthesis of 1, 4-Polybutadiene Dendrimer- Arborescent Polymer Hybrids. *Macromolecules* **2010**, *43* (8), 3672–3681.

(36) Parisi, D.; Buenning, E.; Kalafatakis, N.; Gury, L.; Benicewicz, B. C.; Gauthier, M.; Cloitre, M.; Rubinstein, M.; Kumar, S. K.; Vlassopoulos, D. Universal Polymeric-to-Colloidal Transition in Melts of Hairy Nanoparticles. *ACS nano* **2021**, *15* (10), 16697–16708.

(37) Parisi, D.; Camargo, M.; Makri, K.; Gauthier, M.; Likos, C. N.; Vlassopoulos, D. Effect of Softness on Glass Melting and Re-Entrant Solidification in Mixtures of Soft and Hard Colloids. *The Journal of Chemical Physics* **2021**, *155* (3), 034901.

(38) Daoud, M.; Cotton, J. P. Star Shaped Polymers : A Model for the Conformation and Its Concentration Dependence. *J. Phys. France* **1982**, *43* (3), 531–538.

(39) Milner, S. T.; Lacasse, M.-D.; Graessley, W. W. Why $\chi$ Is Seldom Zero for Polymer-Solvent Mixtures. *Macromolecules* **2009**, *42* (3), 876–886.

(40) Rubinstein, M.; Colby, R. H. *Polymer Physics*; Oxford university press New York, 2003.

(41) Fetters, L. J.; Lohse, D. J.; Colby, R. H. Chain Dimensions and Entanglement Spacings. In *Physical properties of polymers handbook*; Springer, 2007; pp 447–454.

(42) Joanny, J. F.; Leibler, L.; De Gennes, P. G. Effects of Polymer Solutions on Colloid Stability. *Journal of Polymer Science: Polymer Physics Edition* **1979**, *17* (6), 1073–1084.

(43) Truzzolillo, D.; Vlassopoulos, D.; Gauthier, M.; Munam, A. Thermal Melting in Depletion Gels of Hairy Nanoparticles. *Soft Matter* **2013**, *9* (38), 9088–9093.

(44) Weese, J. A Regularization Method for Nonlinear Ill-Posed Problems. *Computer Physics Communications* **1993**, *77* (3), 429–440.

(45) Rogers, S. A.; Vlassopoulos, D.; Callaghan, P. T. Aging, Yielding, and Shear Banding in Soft Colloidal Glasses. *Physical review letters* **2008**, *100* (12), 128304.

(46) Cloitre, M.; Borrega, R.; Leibler, L. Rheological Aging and Rejuvenation in Microgel Pastes. *Physical Review Letters* **2000**, *85* (22), 4819.

(47) Erwin, B. M.; Vlassopoulos, D.; Cloitre, M. Rheological Fingerprinting of an Aging Soft Colloidal Glass. *Journal of rheology* **2010**, *54* (4), 915–939.

(48) Helgeson, M. E.; Wagner, N. J.; Vlassopoulos, D. Viscoelasticity and Shear Melting of Colloidal Star Polymer Glasses. *Journal of Rheology* **2007**, *51* (2), 297–316.

(49) Jones, D. A. R.; Leary, B.; Boger, D. V. The Rheology of a Concentrated Colloidal Suspension of Hard Spheres. *Journal of colloid and interface science* **1991**, *147* (2), 479–495.

(50) Ketz, R. J.; Prud'homme, R. K.; Graessley, W. W. Rheology of Concentrated Microgel Solutions. *Rheologica Acta* **1988**, *27* (5), 531–539.

(51) Siebenbürger, M.; Fuchs, M.; Winter, H.; Ballauff, M. Viscoelasticity and Shear Flow of Concentrated, Noncrystallizing Colloidal Suspensions: Comparison with Mode-Coupling Theory. *Journal of Rheology* **2009**, *53* (3), 707–726.

(52) Vlassopoulos, D.; Fytas, G.; Pakula, T.; Roovers, J. Multiarm Star Polymers Dynamics. *Journal of Physics: Condensed Matter* **2001**, *13* (41), R855.

(53) Flory, P. J. The Configuration of Real Polymer Chains. *The Journal of Chemical Physics* **1949**, *17* (3), 303–310.

(54) Hidalgo-Alvarez, R. *Structure and Functional Properties of Colloidal Systems*; CRC Press, 2009.

(55) Randisi, F.; Pelissetto, A. High-Functionality Star-Branched Macromolecules: Polymer Size and Virial Coefficients. *The Journal of Chemical Physics* **2013**, *139* (15), 154902.

(56) Roovers, J.; Toporowski, P. M.; Douglas, J. Thermodynamic Properties of Dilute and Semidilute Solutions of Regular Star Polymers. *Macromolecules* **1995**, *28* (21), 7064–7070.

(57) Verma, R.; Crocker, J. C.; Lubensky, T. C.; Yodh, A. G. Entropic Colloidal Interactions in Concentrated DNA Solutions. *Physical review letters* **1998**, *81* (18), 4004.

(58) Pham, K. N.; Petekidis, G.; Vlassopoulos, D.; Egelhaaf, S. U.; Poon, W. C. K.; Pusey, P. N. Yielding Behavior of Repulsion-and Attraction-Dominated Colloidal Glasses. *Journal of Rheology* **2008**, *52* (2), 649–676.

(59) Parisi, D.; Ruan, Y.; Ochbaum, G.; Silmore, K. S.; Cullari, L. L.; Liu, C.-Y.; Bitton, R.; Regev, O.; Swan, J. W.; Loppinet, B. Short and Soft: Multidomain Organization, Tunable Dynamics, and Jamming in Suspensions of Grafted Colloidal Cylinders with a Small Aspect Ratio. *Langmuir* **2019**, *35* (52), 17103–17113.

(60) Winter, H. H.; Chambon, F. Analysis of Linear Viscoelasticity of a Crosslinking Polymer at the Gel Point. *Journal of rheology* **1986**, *30* (2), 367–382.

(61) Watanabe, H.; Sato, T.; Osaki, K.; Aoki, Y.; Li, L.; Kakiuchi, M.; Yao, M.-L. Rheological Images of Poly (Vinyl Chloride) Gels. 4. Nonlinear Behavior in a Critical Gel State. *Macromolecules* **1998**, *31* (13), 4198–4204.

(62) Li, L.; Aoki, Y. Rheological Images of Poly (Vinyl Chloride) Gels. 1. The Dependence of Sol- Gel Transition on Concentration. *Macromolecules* **1997**, *30* (25), 7835–7841.

(63) Royall, C. P.; Williams, S. R.; Tanaka, H. Vitrification and Gelation in Sticky Spheres. *The Journal of chemical physics* **2018**, *148* (4), 044501.

(64) Helgeson, M. E.; Wagner, N. J.; Vlassopoulos, D. Viscoelasticity and Shear Melting of Colloidal Star Polymer Glasses. *Journal of Rheology* **2007**, *51* (2), 297–316.

(65) Cloitre, M.; Borrega, R.; Monti, F.; Leibler, L. Glassy Dynamics and Flow Properties of Soft Colloidal Pastes. *Physical review letters* **2003**, *90* (6), 068303.

(66) Mewis, J.; Wagner, N. J. *Colloidal Suspension Rheology*; Cambridge university press, 2012.

(67) Bale, H. D.; Schmidt, P. W. Small-Angle X-Ray-Scattering Investigation of Submicroscopic Porosity with Fractal Properties. *Physical Review Letters* **1984**, *53* (6), 596.

(68) Besselink, R.; Stawski, T. M.; Van Driessche, A. E. S.; Benning, L. G. Not Just Fractal Surfaces, but Surface Fractal Aggregates: Derivation of the Expression for the Structure Factor and Its Applications. *The Journal of chemical physics* **2016**, *145* (21), 211908.

(69) Verhaegh, N. A.; Asnaghi, D.; Lekkerkerker, H. N.; Giglio, M.; Cipelletti, L. Transient Gelation by Spinodal Decomposition in Colloid-Polymer Mixtures. *Physica A: Statistical Mechanics and its Applications* **1997**, *242* (1–2), 104–118.

# Supplementary Material

# Gelation and Re-entrance in Mixtures of Soft Colloids and Linear Polymers of Equal Size

D. Parisi,[1,2,3*] D. Truzzolillo,[4*] A. H. Slim,[5] P. Dieudonné-George,[4] S. Narayanan,[6] J. C. Conrad,[5] V. D. Deepak,[7] M. Gauthier,[7] and D. Vlassopoulos[1,2]

[1]FORTH, Institute of Electronic Structure and Laser, 70013 Heraklion, Crete, Greece

[2]Department of Materials Science and Technology, University of Crete, 70013 Heraklion, Crete, Greece

[3]Department of Chemical Engineering, Product Technology, University of Groningen, Nijenborgh 4, 9747 AG Groningen, The Netherlands

[4]Laboratoire Charles Coulomb (L2C), UMR 5221 CNRS Université de Montpellier, Montpellier, France.

[5]Department of Chemical and Biomolecular Engineering, University of Houston, Houston, Texas 77204-4004, United States

[6]Advanced Photon Source, Argonne National Laboratory, Argonne, IL 60439, United States

[7]Department of Chemistry, University of Waterloo, Waterloo, Ontario N2L 3G1, Canada

*corresponding authors: *domenico.truzzolillo@umontpellier.fr*
　　　　　　　　　　　　*d.parisi@rug.nl*

## Table of Contents

# Dynamic light scattering (DLS) measurements of L243 linear polymer in squalene

DLS measurements were performed using the photon correlation spectroscopy technique[1]. The intensity autocorrelation function $G(q,t) = \langle I(q,t)I(q)\rangle/|I(q)|^2$ is obtained over a broad range of lag times ($10^{-7} - 10^3$ s) at different scattering wavevectors q, with an ALV-5000 goniometer/correlator setup (ALV, Germany) equipped with a Nd:YAG laser at wavelength λ = 532 nm (Oxxius, France). The scattering wavevector is computed as $q = (4\pi n/\lambda)\sin(\theta/2)$ where n and θ are the refractive index of the solvent and the scattering angle, respectively. Measurements were performed in the co-polarized (VV) geometry under homodyne beating conditions. The intermediate scattering function C(q,t) probing concentration fluctuations was extracted from the experimental intensity autocorrelation function acquired with the photomultiplier tube, $G(q,t) : C(q,t) = \left[\dfrac{G(q,t)-1}{A}\right]^{1/2}$, where A ≤ 1 is an instrumental coherence factor, equal to 0.6 for the aforementioned setup.

The experiments were performed under dilute conditions for characterization purposes, and a single exponential relaxation process was observed. By applying the Stokes-Einstein-Sutherland relation $D = \dfrac{k_B T}{6\pi R_H \eta_{squalene}}$, which relates the diffusion coefficient with the hydrodynamic size of the particles, the thermal energy $k_B T$, and the viscosity of the suspending medium $\eta_{squalene}$, the hydrodynamic radii were extracted. All the measurements were performed at 20 °C. Figure S1 below shows the DLS results for sample L243 in squalene.

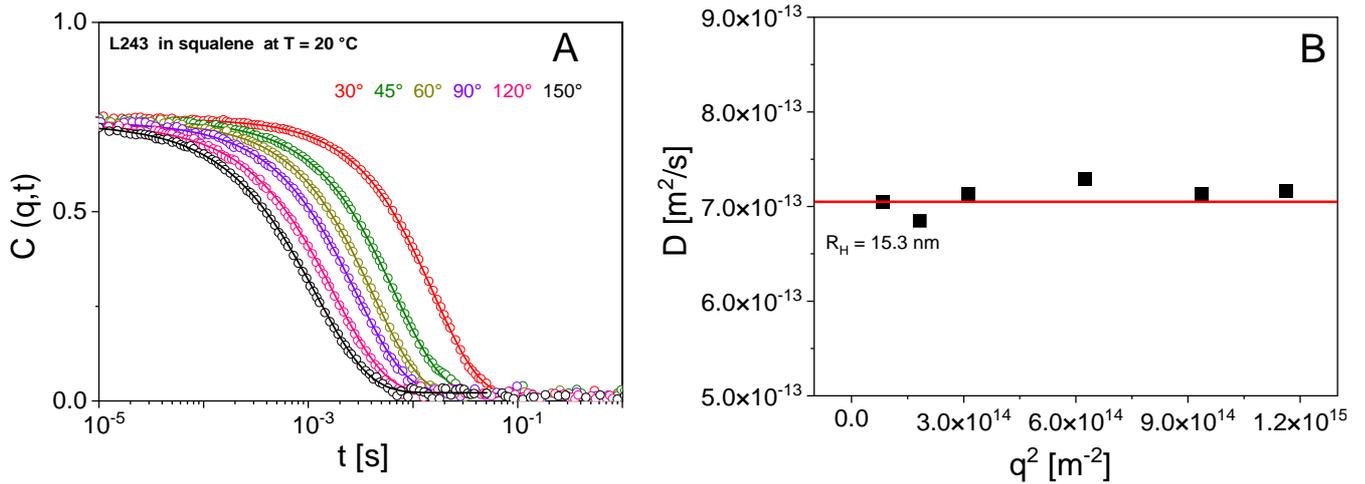

**Figure S1.** (A) Normalized intermediate scattering function C(q,t) for sample L243 in squalene at $3\times10^{-4}$ g/ml. Each color corresponds to a different scattering angle (red: 30°, green: 45°, dark yellow: 60°, purple: 90°, pink: 120°, black: 150°). The solid lines represent fits to single exponential functions. (B) Diffusion coefficient as a function of $q^2$ for the L243 sample. From this plot, the hydrodynamic radius ($R_H$) was extracted using the extrapolated value of D at $q^2 = 0$, and the Stokes-Einstein- Sutherland equation for the diffusion of spherical particles in a medium.

## Linear viscoelasticity: creep conversion, and frequency response of the $\phi_s = 0.5$ and $\phi_s = 0.7$ mixtures

Creep compliance measurements for a representative S362–L1000 mixture at $\phi_s = 0.83$ and $C_L = 7.5$ wt% are provided in Fig. S2A. The creep compliance was always measured at least at two different shear stresses (see Figure S2) to ensure that the measurements were in the linear viscoleastic response region. Deviations within 10% between the data sets were considered acceptable, given the difficulty of such experiments with aging systems. Indeed, the structure of colloid–polymer mixtures may change over time[2], making the creep experiments more difficult to reproduce. Note that the first two time decades, where the known "ringing" phenomenon occurs[3], are usually not considered for creep data conversion. Nevertheless, the conversion of this region is still shown for completeness. Figure S2B shows conversion of the creep compliance data into dynamic moduli as a function of the oscillation frequency. Dynamic moduli obtained from small amplitude oscillatory shear measurements are also shown in the same Figure as symbols, to validate the creep data and their conversion.

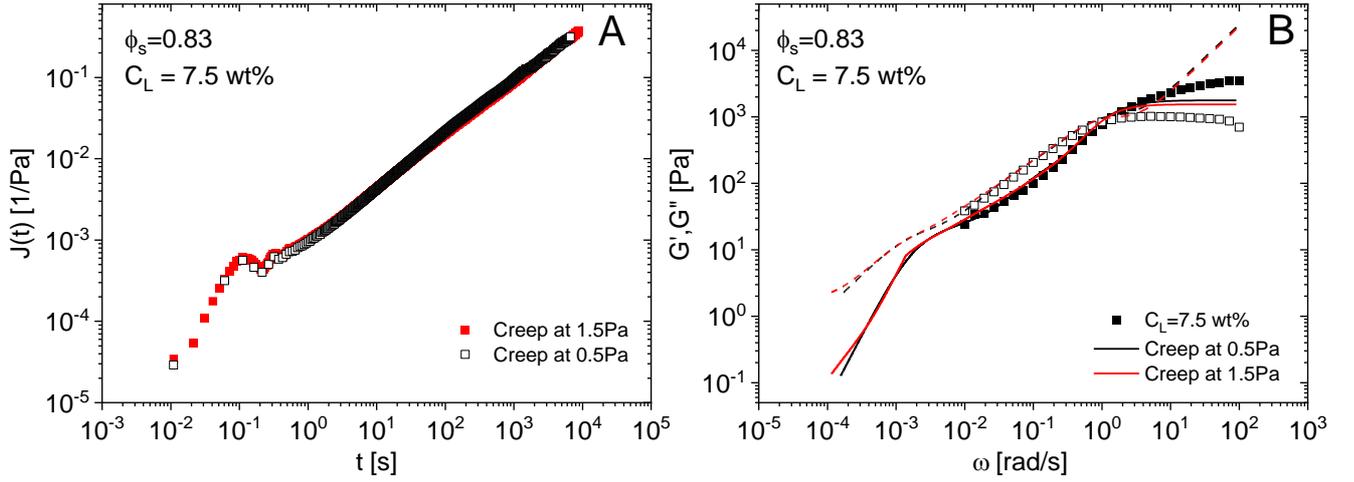

**Figure S2.** (A) Creep compliance as a function of time for a S362–L1000 mixture at $\phi_s = 0.83$ and $C_L = 7.5$ wt%, measured at two different shear stresses: 1 Pa (blue symbols) and 1.5 Pa (black symbols). (B) Storage G' (solid symbols) and loss G" (open symbols) moduli as a function of the oscillatory frequency ω for a S362–L1000 mixture at $\phi_s = 0.83$ and $C_L = 7.5$ wt%. The data obtained from small amplitude oscillatory shear are reported as symbols, whereas the lines represent the creep conversion into storage modulus (solid line) and loss modulus (dashed line) at 1 Pa (black lines) and 1.5 Pa (red lines). Experiments performed at 20 °C.

A short description of the creep conversion follows in order. The shear creep compliance J(t) should be first converted into the continuous spectrum of retardation times, L(τ). J.D. Ferry proposed the exact interrelation between the J(t) and L(τ) functions[4]:

$$J(t) = J_g + \int_{-\infty}^{\infty} L\left(1 - e^{-\frac{t}{\tau}}\right) d\ln(\tau) + t/\eta_0 \qquad (S1)$$

with $J_g$ being an instantaneous compliance added to allow for the possibility of a discrete contribution with t = 0. $\eta_0$ is the zero-shear viscosity. The retardation spectrum enables the determination of the dynamic compliance moduli:

$$J'(\omega) = J_g + \int_{-\infty}^{\infty} \left[\frac{L}{(1 + \omega^2\tau^2)}\right] d\ln(\tau) \qquad (S2)$$

$$J''(\omega) = \int_{-\infty}^{\infty} \left[\frac{L\omega\tau}{(1 + \omega^2\tau^2)}\right] d\ln(\tau) + 1/(\omega\eta_0) \qquad (S3)$$

The latter can be directly converted into dynamic moduli by means of:

$$J'(\omega) = \frac{G'(\omega)}{G'^2(\omega) + G''^2(\omega)} \qquad (S4)$$

$$J''(\omega) = \frac{G''(\omega)}{G'^2(\omega) + G''^2(\omega)} \qquad (S5)$$

The challenge is then the conversion between the shear creep compliance and the retardation spectrum, and this into dynamic compliance. Determining L(τ) from J(t) represents a mathematically ill-posed problem, as various L(τ) functions can lead to the same J(t). The most renowned mathematical method is the regularization method of Tikhonov[5]. In the present case, the relation between the physically interesting function and the experimental data is given by nonlinear integral equations, further complicating the math. Weese proposed the so-called nonlinear regularization method[6]; a mathematical treatment that allows for the determination of a physically interesting function, described by a nonlinear integration function. Weese implemented his model in a software named NLREG. In our case, we implemented the mathematical treatment in Mathematica®. This method is quite accurate, and the only significant source of error is the experimental error in measuring J(t). Figure S3 below shows the retardation spectrum at the two different stresses shown in Figure S2A, including error bars (within the symbol size).

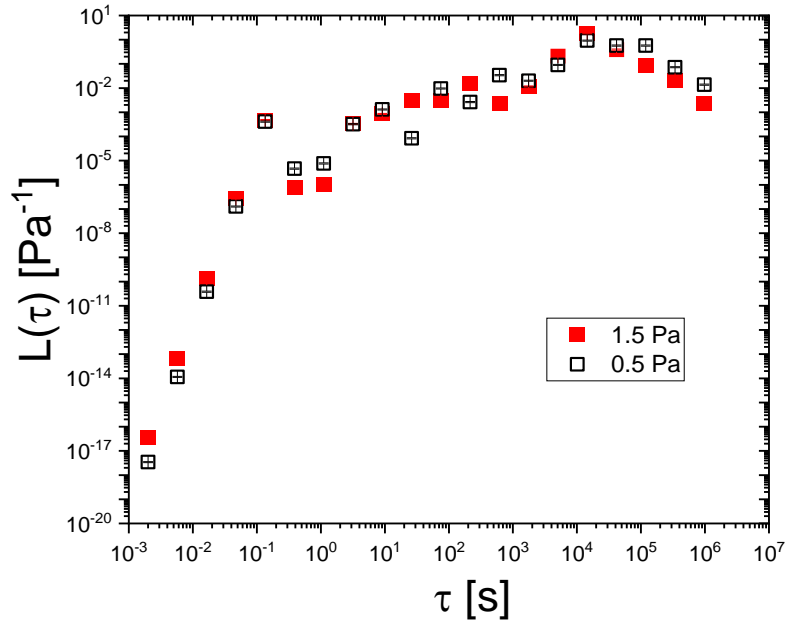

**Figure S3.** Retardation spectrum of the data shown in Figure S2A, obtained from the method proposed by Weese[6].

Fig. S4 depicts the linear viscoelastic spectra in terms of storage (G') and loss (G") moduli as a function of the oscillatory frequency for the S362–L1000 mixture at fixed $\phi_s$ = 0.5 (left panel) and $\phi_s$ = 0.7 (right panel) for increasing $C_L$. Note that, in contrast to the mixtures reported in the main text, no gelation was observed at any $C_L$ value, albeit the colloidal mode was still observable at low frequencies.

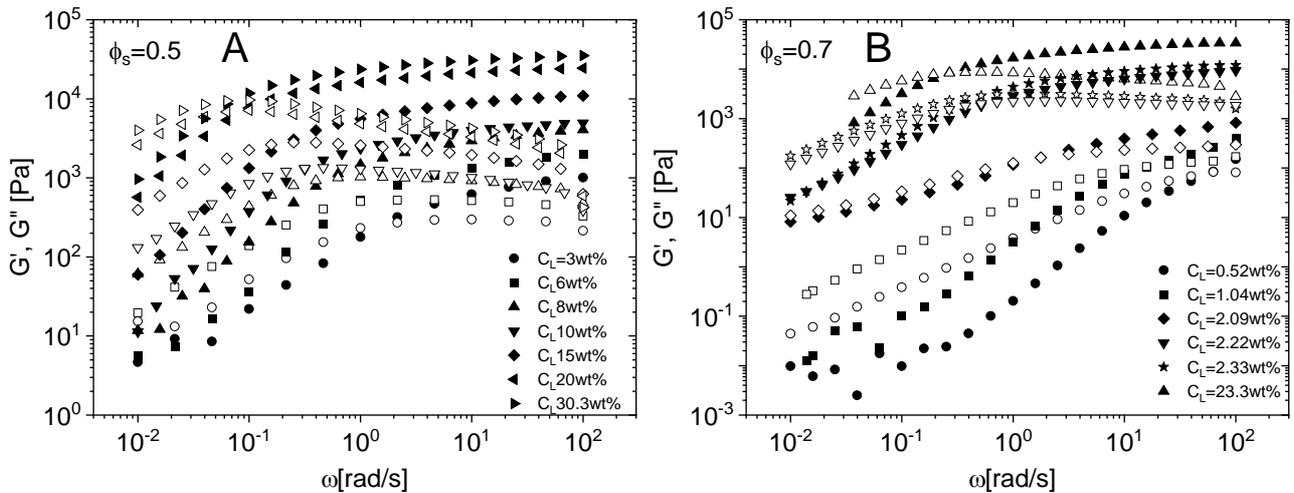

**Figure S4.** G' (solid symbols) and G" (open symbols) as a function of $\omega$ for the S362–L1000 mixture at fixed $\phi_s$ = 0.5 (panel A) and $\phi_s$ = 0.7 (panel B) for increasing $C_L$. Experiments performed at 20 °C.

## On osmotic shrinkage: Flory-type approach

By using Flory-type arguments concerning the size of a star in a bath of linear polymer chains[7–10], it is possible to estimate the shrinkage of the star and the gain in free volume due to the osmotic pressure exerted by the linear chains onto the star. Firstly, the shrinkage of the star was calculated through the osmotic theory[7–10]. The free energy cost of inserting a star polymer with radius R in a solution of homopolymer linear chains corresponds to the mechanical work that needs to be done to create a spherical cavity within the solution:

$$\beta F_{os}(R) = \frac{4\pi}{3} R^3 a^3 \beta \Pi(\phi_L) \tag{S6}$$

where $\beta = 1/k_B T$, $a$ is the monomer radius (0.5 nm[11,12]), and $\Pi(\phi_L)$ the osmotic pressure exerted by the polymer matrix and is written as

$$\beta \Pi(\phi_L) = C_L [1 + P(\phi_L)] \tag{S7}$$

with

$$P(x) = \frac{1}{2} x \, exp\left\{\frac{1}{4}\left[\frac{1}{x} + 1 - \frac{1}{x}\ln(x+1)\right]\right\} \tag{S8}$$

and

$$x(\phi_L) = \frac{1}{2} \phi_L \pi^2 \left[1 - \frac{1}{4}\left(\ln 2 + \frac{1}{2}\right)\right] \tag{S9}$$

where R represents the radius of the spherical cavity, which does not necessarily coincide with the size of the star $R_s$ because of the penetrability of the stars. Camargo and Likos[13] have shown that a chain can penetrate a star up to a distance $\sigma = 4/3 R_s = R$, which is the corona size of the star. It follows that $R = bR_s = 1.3 R_s$. This numerical prefactor is essentially independent of the star/linear ratio, provided that it remains larger than 1. However, as in the present case, when the stars and linear chains are comparable in size, b must attain larger values because the chains would rather surround the stars than penetrate them. In the present work b = 1.7 was used, as suggested by Wilk *et al.*[8] when stars and linear chains have a comparable size.

The osmotic pressure is not the only contribution to the free energy. Associated with a star of radius $R_s$ of functionality $f$ and degree of polymerization $N_s$, the elastic and interaction free energies (excluded volume) are respectively given by

$$\beta F_{el}(R) = \frac{3}{2}\frac{afR_s^2}{N_s} \quad (S10)$$

and

$$\beta F_{int}(R) = \frac{a^3 v(fN_s)^2}{2R_s^3} \quad (S11)$$

where v is the excluded volume parameter in reduced units, which has been given the value 1, corresponding to good solvency conditions[7,8]. The extent of shrinkage of the stars due to the presence of linear chains is determined as the value at which the overall free energy reaches a minimum. Thus, minimizing the free energy with respect to $R_s$ the following result is obtained:

$$\frac{3afR_s}{N_s} + 4\pi b^3 R_s^2 a^3 C_L[1 + P(\phi_L)] - \frac{3a^3 v(fN_s)^2}{2R_s^4} = 0 \quad (S12)$$

By defining the overlap concentration as $C_L^* = a^{-3}N_L^{-3v}$[7,8] with $v = 3/5$ as the Flory exponent and $N_L$ representing the degree of polymerization of the linear chains, and using the scaling relation for the size ratio between linear chains and stars ($\delta$) as[10] $\delta = \left(\frac{1}{f^{1/5}}\right)\left(\frac{N_L}{N_s}\right)^v$, equation S12 becomes

$$\frac{3afR_s}{N_s} + 4\pi b^3 R_s^2 \delta^{-3} N_s^{-3v} f^{-3/5} \phi_L[1 + P(\phi_L)] - \frac{3a^3 v(fN_s)^2}{2R_s^4} = 0 \quad (S13)$$

The effective volume fraction of chains and the size of stars can be self-consistently evaluated as described in Ref. 7 for different star concentrations.

Results are reported as the shrinkage factor $g(\phi_L) = R_s(\phi_L)/R_{H,0}$ against $\phi_L$ with $f = 362$, $N_s = 1240$, $N_L = 19600$ and $\delta = 5$ in Figure S5. $R_{H,0}$ is the hydrodynamic radius of the star in the absence of linear chains. Along with the theoretical predictions, are also reported the calculated shrinkage values (see Ref. 14) and the spacing of the scatterers determined from the SAXS measurements (see main text). Good agreement was observed between the calculated values, theoretical predictions and the SAXS data, further corroborating the osmotic shrinkage effect of the stars exerted by the star themselves and the linear polymer chains.

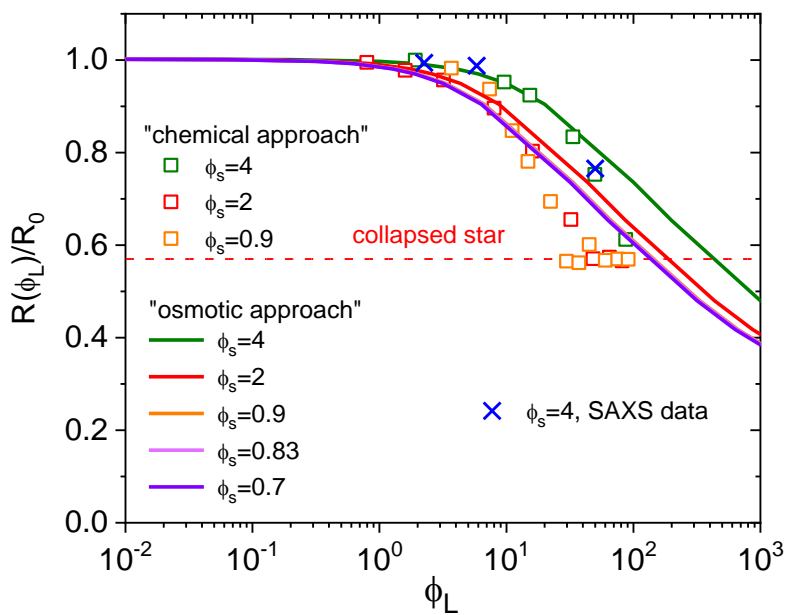

**Figure S5.** Osmotic shrinkage expressed as the ratio between the radius of the stars $R(\phi_L)$ at non-zero $\phi_L$ and its value $R_0$ in absence of chains as a function of the volume fraction of linear polymer chains $\phi_L$. The lines represent the theoretical prediction, whereas the square symbols are the calculated values discussed in Ref. 14 (chemical approach). Predictions for $\phi_s$=0.7, 0.83, 0.9 do not show any difference within the numerical precision of our code. The cross symbols were extracted from the SAXS data (see Figure 5 in the main text and relevant discussion). The dashed red line represents the calculated size of the fully shrunken star, which can be seen as the radius of a sphere containing $fN_a$ close-packed monomers, where $f$ is the star functionality and $N_a$ the Kuhn degree of polymerization of one arm[14].

Table S1 below provides a comparison of the star radius for the S362 suspensions at various $\phi_s$ by two different approaches: 1) chemical approach[14], 2) and SAXS measurements.

Table S1. S362 estimated radius at various $\phi_s$.

| $\phi_s$ | chemical approach[14] | $d_{cc}/2 = R = \pi a/q_{max}$ |
|---|---|---|
| 0.9 | 38.3 | |
| 2 | 32.4 | 33.0 |
| 3 | | 27.0 |
| 4 | 27.7 | 24.8 |
| 5 | | 23.3 |

# References

(1) Berne, B. J.; Pecora, R. *Dynamic Light Scattering: With Applications to Chemistry, Biology, and Physics*; Courier Corporation, **2000**.

(2) Truzzolillo, D.; Vlassopoulos, D.; Munam, A.; Gauthier, M. Depletion Gels from Dense Soft Colloids: Rheology and Thermoreversible Melting. *Journal of Rheology* **2014**, *58* (5), 1441–1462.

(3) Ewoldt, R. H.; McKinley, G. H. Creep Ringing in Rheometry or How to Deal with Oft-Discarded Data in Step Stress Tests! *Rheol. Bull* **2007**, *76* (4).

(4) Ferry, J. D. *Viscoelastic Properties of Polymers*; John Wiley & Sons, **1980**.

(5) Nashed, M. Z. The Theory of Tikhonov Regularization for Fredholm Equations of the First Kind (Cw Groetsch). *Siam Review* **1986**, *28* (1), 116–118.

(6) Weese, J. A Regularization Method for Nonlinear Ill-Posed Problems. *Computer Physics Communications* **1993**, *77* (3), 429–440.

(7) Truzzolillo, D.; Vlassopoulos, D.; Gauthier, M. Osmotic Interactions, Rheology, and Arrested Phase Separation of Star–Linear Polymer Mixtures. *Macromolecules* **2011**, *44* (12), 5043–5052.

(8) Wilk, A.; Huißmann, S.; Stiakakis, E.; Kohlbrecher, J.; Vlassopoulos, D.; Likos, C. N.; Meier, G.; Dhont, J. K. G.; Petekidis, G.; Vavrin, R. Osmotic Shrinkage in Star/Linear Polymer Mixtures. *The European Physical Journal E* **2010**, *32* (2), 127–134.

(9) Flory, P. J. The Configuration of Real Polymer Chains. *The Journal of Chemical Physics* **1949**, *17* (3), 303–310.

(10) Likos, C. N. Effective Interactions in Soft Condensed Matter Physics. *Physics Reports* **2001**, *348* (4–5), 267–439.

(11) Likos, C. N.; Löwen, H.; Poppe, A.; Willner, L.; Roovers, J.; Cubitt, B.; Richter, D. Ordering Phenomena of Star Polymer Solutions Approaching the Θ State. *Physical Review E* **1998**, *58* (5), 6299–6307.

(12) Li, X.; Ma, X.; Huang, L.; Liang, H. Developing Coarse-Grained Force Fields for Cis-Poly (1, 4-Butadiene) from the Atomistic Simulation. *Polymer* **2005**, *46* (17), 6507–6512.

(13) Camargo, M.; Likos, C. N. Unusual Features of Depletion Interactions in Soft Polymer-Based Colloids Mixed with Linear Homopolymers. *Physical review letters* **2010**, *104* (7), 078301.

(14) Parisi, D.; Truzzolillo, D.; Deepak, V. D.; Gauthier, M.; Vlassopoulos, D. Transition from Confined to Bulk Dynamics in Symmetric Star–Linear Polymer Mixtures. *Macromolecules* **2019**, *52* (15), 5872–5883.